\documentclass{appolb}
\usepackage{epsfig}

\usepackage{amsfonts}
\usepackage{amssymb}
\usepackage{amsmath}

\usepackage{ulem}
\usepackage{color}

\renewcommand\sout{\bgroup \color{blue} \ULdepth=-.5ex \ULset}

\def\Slash#1{\not\!\!#1}

\begin{document}
\title{Matter-Antimatter Coexistence Method for Finite Density QCD 
\thanks{Presented at the International Workshop on ``Excited QCD 2017'' 
(eQCD 2017), May 7-13, 2017, Sintra, Portugal}%
}
\author{Hideo Suganuma
\address{Department of Physics, Graduate School of Science, Kyoto University, 
\\
Kitashirakawaoiwake, Sakyo, Kyoto 606-8502, Japan}
}
\maketitle
\begin{abstract}
We propose a ``matter-antimatter coexistence method'' for finite-density lattice QCD, 
aiming at a possible solution of the sign problem. 
In this method, we consider matter and anti-matter systems 
on two parallel ${\bf R}^4$-sheets in five-dimensional Euclidean space-time.
For the matter system $M$ with a chemical potential $\mu \in {\bf C}$ 
on a ${\bf R}^4$-sheet, 
we also prepare the anti-matter system $\bar M$ with $-\mu^*$ 
on the other ${\bf R}^4$-sheet shifted in the fifth direction. 
In the lattice QCD formalism, 
we introduce a correlation term between the gauge variables 
$U_\nu \equiv e^{iagA_\nu}$ in $M$ and 
$\tilde U_\nu \equiv e^{iag \tilde A_\nu}$ in $\bar M$, such as 
$S_\lambda \equiv 
\sum_{x,\nu} 2\lambda  \{N_c-{\rm Re~tr} [U_\nu(x) \tilde U_\nu^\dagger(x)]\} 
\simeq \sum_x \frac{1}{2}\lambda a^2 \{A_\nu^a(x)-\tilde A_\nu^a(x)\}^2$  
with a real parameter $\lambda$. 
In the limit of $\lambda \rightarrow \infty$, 
a strong constraint $\tilde U_\nu(x)=U_\nu(x)$ is realized, 
and the total fermionic determinant is real and non-negative. 
In the limit of $\lambda \rightarrow 0$, this system goes to 
two separated ordinary QCD systems 
with the chemical potential of $\mu$ and $-\mu^*$. 
On a finite-volume lattice, 
if one takes an enough large value of $\lambda$, 
$\tilde U_\nu(x) \simeq U_\nu(x)$ is realized and there occurs  
a phase cancellation approximately 
between two fermionic determinants in $M$ and $\bar M$, 
which is expected to suppress the sign problem 
and to make the lattice calculation possible.  
For the obtained gauge configurations of the coexistence system, 
matter-side quantities are evaluated 
through their measurement only for the matter part $M$. 
By the calculations with gradually decreasing $\lambda$ and 
their extrapolation to $\lambda=0$, 
physical quantities in finite density QCD are expected to be estimated.
\end{abstract}

\def\slash#1{\not\!#1}
\def\slashb#1{\not\!\!#1}
\def\slashbb#1{\not\!\!\!#1}

\section{Introduction}

The lattice QCD Monte Carlo calculation has revealed many aspects of 
the QCD vacuum and hadron properties in both zero and finite temperatures.
At finite density, however, lattice QCD is not yet well investigated, 
because of a serious problem called the ``sign problem'' \cite{P83,FK02}, 
which originates from  the complex value including minus sign 
of the QCD action and the fermionic determinant at finite density, 
even in the Euclidean metric \cite{HK84}.
In fact, the Euclidean QCD action $S[A, \psi, \bar \psi;\mu]$ 
at finite density with the chemical potential $\mu$ is generally complex, 
\begin{eqnarray}
S[A, \psi, \bar \psi;\mu]=S_G[A]
+\int d^4x \{\bar \psi (\Slash D+m+\mu\gamma_4) \psi \} \in {\bf C},
\end{eqnarray}
with the gauge action $S_G[A] \in {\bf R}$ and 
covariant derivative $D^\nu \equiv \partial^\nu +igA^\nu$. 
Then, the action factor cannot be identified as a probability density 
in the QCD generating functional, unlike ordinary lattice QCD calculations. 

In this paper, aiming at a possible solution of the sign problem, 
we propose a new approach of a ``matter-antimatter coexistence method''  
for lattice QCD at finite density with a general chemical potential $\mu \in {\bf C}$.

\section{Matter-Antimatter Coexistence Method}

Our strategy is to use a cancelation of the phase factors of 
the fermionic determinants between 
a matter system with $\mu$ and an anti-matter system with $-\mu^*$, 
and our method is based on the general property \cite{HK84},
\begin{eqnarray}
S[A, \psi, \bar \psi; \mu]^*=S[A, \psi, \bar \psi; -\mu^*],
\label{eq:action}
\end{eqnarray}
for the Euclidean QCD action $S[A, \psi, \bar \psi; \mu]$
in the presence of the chemical potential $\mu \in {\bf C}$.
Actually, the fermionic kernel $D_F$ corresponding to $\Slash{D}+m$ 
generally satisfies  
$D_F^\dagger=\gamma_5 D_F \gamma_5$ in lattice QCD, 
so that one finds 
\begin{eqnarray}
[\bar \psi (D_F+\mu \gamma_4) \psi ]^*
=\bar \psi (D_F-\mu^*\gamma_4) \psi,
\end{eqnarray}
which leads to the relation (\ref{eq:action}), and 
\begin{eqnarray}
{\rm Det} (D_F+\mu \gamma_4)^*
={\rm Det} (D_F-\mu^*\gamma_4).
\label{eq:det}
\end{eqnarray}

\subsection{Definition and Setup}

In the ``matter-antimatter coexistence method'', 
we consider matter and anti-matter systems 
on two parallel ${\bf R}^4$-sheets in five-dimensional Euclidean space-time.
For the matter system $M$ with a chemical potential $\mu \in {\bf C}$ 
on a ${\bf R}^4$-sheet, we also prepare the anti-matter system $\bar M$ with $-\mu^*$ 
on the other ${\bf R}^4$-sheet shifted in the fifth direction, as shown in Fig.~1.

\begin{figure}[h]
\begin{center}
\includegraphics[scale=0.72]{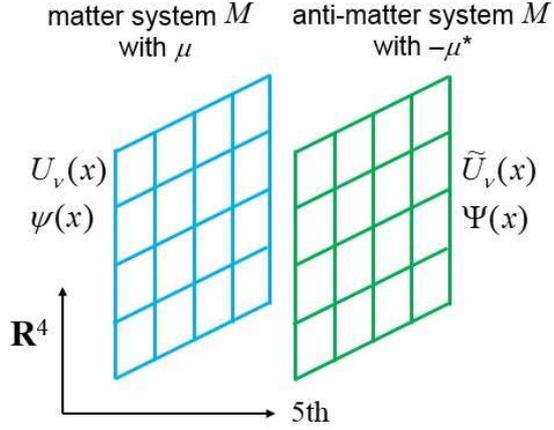}
\caption{
The matter-antimatter coexistence system in five-dimensional Euclidean space-time.
We put the matter system $M$ with $\mu$, $U_\nu(x)$ and $\psi(x)$  
on a ${\bf R}^4$-sheet, and the anti-matter system $\bar M$ 
with $-\mu^*$, $\tilde U_\nu(x)=U_\nu(x+\hat 5)$ and $\Psi(x)=\psi(x+\hat 5)$ 
on the other ${\bf R}^4$-sheet shifted in the fifth direction.
}
\end{center}
\end{figure}

We put an ordinary fermion field $\psi(x)$ with the mass $m$ 
and the gauge variable $U_\mu (x) \equiv e^{iag A_\nu(x)}$ 
at $x \in {\bf R}^4$ on the matter system $M$, 
and we put the other fermion field $\Psi(x) \equiv \psi(x+\hat 5)$ 
with the same mass $m$ and the gauge variable 
$\tilde U_\mu(x) \equiv e^{iag \tilde A_\nu(x)} \equiv U_\mu (x+\hat 5)$ 
on the anti-matter system $\bar M$.

In the lattice QCD formalism, 
we introduce a correlation term between the gauge variables $U_\nu(x)$ in $M$ 
and $\tilde U_\nu(x)$ in $\bar M$ at $x \in {\bf R}^4$, such as 
\begin{eqnarray}
S_\lambda \equiv \sum_{x,\nu} 2\lambda 
\{N_c-{\rm Re~tr} [U_\nu(x) \tilde U_\nu^\dagger(x)]\}
\label{eq:corr}
\end{eqnarray}
with a real parameter $\lambda$ ($\ge 0$), 
which connects two different situations: 
$\tilde U_\nu(x)=U_\nu(x)$ in $\lambda \rightarrow \infty$ 
and two separated QCD systems in $\lambda \rightarrow 0$.
Near the continuum limit, this additional term becomes 
\begin{eqnarray}
S_\lambda \simeq \sum_x \frac{1}{2}\lambda a^2 \{A_\nu^a(x)-\tilde A_\nu^a(x)\}^2
\simeq \int d^4x~\frac{1}{2} \lambda_{\rm phys} \{A_\nu^a(x)-\tilde A_\nu^a(x)\}^2
\end{eqnarray}
with $\lambda_{\rm phys} \equiv \lambda a^{-2}$.

In fact, the total lattice action 
in this method is written as
\begin{eqnarray}
S&=&S_G[U]+\sum_x \bar \psi(D_F[U]+\mu\gamma_4)\psi 
+S_G[\tilde U]+\sum_x \bar \Psi(D_F[\tilde U]-\mu^* \gamma_4)\Psi \cr
&+&\sum_{x,\nu} 
2\lambda \{N_c-{\rm Re~tr} [U_\nu(x) \tilde U_\nu^\dagger(x)]\}
\end{eqnarray}
with the gauge action $S_G[U] \in {\bf R}$ 
and the fermionic kernel $D_F[U]$ in lattice QCD.
After integrating out the fermion fields $\psi$ and $\Psi$, 
the generating functional of this theory reads 
\begin{eqnarray}
Z&=&\int DU e^{-S_G[U]}{\rm Det} (D_F[U]+\mu\gamma_4) 
\int D{\tilde U} e^{-S_G[\tilde U]}{\rm Det} (D_F[\tilde U]-\mu^* \gamma_4) \cr
&& e^{-\sum_{x,\nu} 2\lambda \{N_c-{\rm Re~tr} [U_\nu(x) \tilde U_\nu^\dagger(x)]\}} \cr
&=&\int DU \int D{\tilde U} e^{-(S_G[U]+S_G[\tilde U])}
{\rm Det} \{(D_F[U]+\mu\gamma_4) (D_F[\tilde U]-\mu^* \gamma_4)\} \cr
&& e^{-\sum_{x,\nu} 2\lambda \{N_c-{\rm Re~tr} [U_\nu(x) \tilde U_\nu^\dagger(x)]\}}.
\label{eq:gene}
\end{eqnarray}
In the continuum limit, this generating functional is expressed as
\begin{eqnarray}
Z_{\rm cont}&=&\int DA \int D{\tilde A}  e^{-(S_G[A]+S_G[\tilde A])}
{\rm Det} \{(\Slash{D}+m+\mu\gamma_4) (\Slash{\tilde D}+m-\mu^* \gamma_4)\} \cr
&& e^{-\int d^4x \frac{1}{2}\lambda_{\rm phys} \{A_\nu^a(x)-\tilde A_\nu^a(x)\}^2}
\end{eqnarray}
with the continuum gauge action $S_G[A] \in {\bf R}$ and 
$\tilde D^\nu \equiv \partial^\nu+i g\tilde A^\nu$. 

In the practical lattice calculation with the Monte Carlo method, 
the fermionic determinant in $Z$  
is factorized into its amplitude and phase factor as
\begin{eqnarray}
Z&=&\int DU \int D{\tilde U} e^{-(S_G[U]+S_G[\tilde U])}
e^{-\sum_{x,\nu} 2\lambda \{N_c-{\rm Re~tr} [U_\nu(x) \tilde U_\nu^\dagger(x)]\}} \cr
&& |{\rm Det} \{(D_F[U]+\mu\gamma_4) (D_F[\tilde U]-\mu^* \gamma_4)\}| \cdot 
O_{\rm phase}[U,\tilde U],
\end{eqnarray}
and the phase factor of the total fermionic determinant  
\begin{eqnarray}
O_{\rm phase}[U,\tilde U]
\equiv e^{i{\rm arg}[{\rm Det} \{(D_F[U]+\mu\gamma_4) (D_F[\tilde U]-\mu^* \gamma_4)\}]}\in {\bf C}
\label{eq:phase}
\end{eqnarray}
is treated as an ``operator'' instead of a probability factor, 
while all other real non-negative factors in $Z$ are 
treated as the probability density.

\subsection{Property and Procedure}

The additional term $S_\lambda$ connects the following 
two different situations as the two limits of the parameter $\lambda$.

\begin{enumerate}
\item 
In the limit of $\lambda \rightarrow \infty$, 
a strong constraint $\tilde U_\nu(x)=U_\nu(x)$ is realized, 
and the phase factors of two fermionic determinants 
${\rm Det} (D_F[U]+\mu\gamma_4)$ and 
${\rm Det} (D_F[\tilde U]-\mu^* \gamma_4)$
are completely cancelled, owing to Eq.~(\ref{eq:det}).
Therefore, the total fermionic determinant is real and non-negative,
\begin{eqnarray}
{\rm Det} \{(D_F[U]+\mu\gamma_4) (D_F[\tilde U=U]-\mu^* \gamma_4)\} \ge 0,
\end{eqnarray} 
and the sign problem is absent \cite{S17}. 
Note however that this system resembles QCD 
with an isospin chemical potential \cite{SS01}.
\item
In the limit of $\lambda \rightarrow 0$, this system goes to 
two separated ordinary QCD systems 
with the chemical potential of $\mu$ and $-\mu^*$, 
although the cancellation of the phase factors cannot be expected between 
the two fermionic determinants 
${\rm Det} (D_F[U]+\mu\gamma_4)$ and 
${\rm Det} (D_F[\tilde U]-\mu^* \gamma_4)$ 
for significantly different $U_\nu(x)$ and $\tilde U_\nu(x)$, which are 
independently generated in the Monte Carlo simulation.
\end{enumerate}

On a four-dimensional finite-volume lattice, 
if an enough large value of $\lambda$ is taken, 
$\tilde U_\nu(x) \simeq U_\nu(x)$ is realized and there occurs  
the phase cancellation approximately 
between the two fermionic determinants 
${\rm Det} (D_F[U]+\mu\gamma_4)$ and 
${\rm Det} (D_F[\tilde U]-\mu^* \gamma_4)$ in $M$ and $\bar M$, 
so that one expects a modest behavior of the phase factor 
$O_{\rm phase}[U,\tilde U]$ in Eq.(\ref{eq:phase}), 
which leads to feasibility of the numerical lattice calculation 
with suppression of the sign problem.

Once the lattice gauge configurations of the coexistence system are obtained 
with the most importance sampling in the Monte Carlo simulation, 
matter-side quantities can be evaluated 
through their measurement only for the matter part $M$ with $\mu$. 

By performing the lattice calculations with gradually decreasing $\lambda$ 
and their extrapolation to $\lambda=0$, 
we expect to estimate the physical quantities in finite density QCD 
with the chemical potential $\mu$.

\section{Summary, Discussion and Outlook}

We have proposed a ``matter-antimatter coexistence method'' 
for the lattice calculation of finite density QCD. 
In this method, we have prepared matter $M$ with $\mu$ 
and anti-matter $\bar M$ with $-\mu^*$  
on two parallel ${\bf R}^4$-sheets in five-dimensional Euclidean space-time, 
and have introduced a correlation term 
$S_\lambda\equiv \sum_{x,\nu} 
2\lambda \{N_c-{\rm Re~tr} [U_\nu(x) \tilde U_\nu^\dagger(x)]\} 
\simeq \sum_x \frac{1}{2}\lambda a^2\{A_\nu^a(x)-\tilde A_\nu^a(x)\}^2$ 
between the gauge variables 
$U_\nu=e^{iag A_\nu}$ in $M$ and $\tilde U_\nu=e^{iag \tilde A_\nu}$ in $\bar M$. 
In the limit of $\lambda \rightarrow \infty$, owing to $\tilde U_\nu(x)=U_\nu(x)$, 
the total fermionic determinant is real and non-negative, 
and the sign problem is absent. 
In the limit of $\lambda \rightarrow 0$, this system goes to 
two separated ordinary QCD systems 
with the chemical potential of $\mu$ and $-\mu^*$.

For an enough large value of $\lambda$, 
$\tilde U_\nu(x) \simeq U_\nu(x)$ is realized and 
a phase cancellation approximately occurs 
between two fermionic determinants in $M$ and $\bar M$, 
which is expected to suppress the sign problem 
and to make the lattice calculation possible. 
For the obtained gauge configurations of the coexistence system, 
matter-side quantities can be evaluated 
by their measurement only for the matter part $M$.
By gradually reducing $\lambda$ and the extrapolation to $\lambda=0$, 
it is expected to obtain estimation of the physical quantities 
in finite density QCD with $\mu$.

In this paper, we have demonstrated this method 
with taking $S_\lambda$ in Eq.(\ref{eq:corr}) 
as the simplest correlation between 
$U_\nu(x)$ in $M$ and $\tilde U_\nu(x)$ in $\bar M$. 
In this method, however, 
there is some variety on the choice of the correlation between 
$U_\nu(x)$ and $\tilde U_\nu(x)$. 
For instance, it may be interesting to consider the other correlation term like 
\begin{eqnarray}
\bar S_\xi &\equiv& \sum_{x} 8\xi
\left( \sum_\nu \{N_c-{\rm Re~tr} [U_\nu(x) \tilde U_\nu^\dagger(x)]\} \right)^3 \cr
&\simeq& \int d^4x~\frac{1}{8} a^2 \xi [\{A_\nu^a(x)-\tilde A_\nu^a(x)\}^2]^3
\end{eqnarray}
with a dimensionless non-negative real parameter $\xi$.
At the classical level, this correlation is an irrelevant interaction 
and it gives vanishing contributions in the continuum limit $a \rightarrow 0$, 
like the Wilson term $-\frac{1}{2} ar \bar \psi D^2\psi$.

The next step is to perform the actual lattice QCD calculation 
at finite density using this method. 
It would be useful to combine this method with the other known ways 
such as the hopping parameter expansion \cite{ASSS14}, 
the complex Langevin method \cite{P83} and the reweighting technique \cite{FK02}.
For example, if one utilizes the hopping parameter expansion, 
the quenched-level analysis becomes possible in this method, 
since the additional term $S_\lambda$ only includes gauge variables.

Efficiency of this method would strongly depend on 
the system parameters, such as the space-time volume $V$, 
the quark mass $m$, the temperature $T$ and the chemical potential $\mu$. 
In any case, this method is expected to enlarge calculable area 
of the QCD phase diagram on $(T, \mu, m, V)$.

\section*{Acknowledgements}
The author would like to thank Prof. P. Bicudo for his useful comment 
and his hospitality during the Workshop.
The author is supported in part by the Grant for Scientific Research 
[(C)15K05076] from the Ministry of Education, Science and Technology of Japan.

\end{document}